\documentclass[onecolumn]{article}

\usepackage{graphicx,epsfig}
\usepackage{times}
\usepackage{hyperref}
\usepackage{amssymb, amsthm}
\usepackage {amssymb,latexsym,epic,eepic,stmaryrd,url}
\usepackage{bbm}
\newtheorem{thm}{Theorem}
\newtheorem{clm}{Claim}
\newtheorem{fact}{Fact}
\newtheorem{lem}{Lemma}
\newtheorem{lem1}{Lemma}

\begin{document}
\title{\Large \bf Acyclic Edge Coloring of Planar Graphs}

\author{ Manu Basavaraju\thanks{{\bf (Corresponding Author) } Computer Science and Automation department,
Indian Institute of Science,
Bangalore- 560012,
India.  {\tt manu@csa.iisc.ernet.in; iammanu@gmail.com}} \and L. Sunil Chandran\thanks{Computer Science and Automation department,
Indian Institute of Science,
Bangalore- 560012,
India.  {\tt sunil@csa.iisc.ernet.in}}
}

\date{}
\pagestyle{plain}
\maketitle

\begin{abstract}

An $acyclic$ edge coloring of a graph is a proper edge coloring such that there are no bichromatic cycles. The \emph{acyclic chromatic index} of a graph is the minimum number k such that there is an acyclic edge coloring using k colors and is denoted by $a'(G)$.  It was conjectured by Alon, Sudakov and Zaks (and much earlier by Fiamcik) that $a'(G)\le \Delta+2$, where $\Delta =\Delta(G)$ denotes the maximum degree of the graph. We prove that if $G$ is a planar graph with maximum degree $\Delta$, then $a'(G)\le \Delta + 12$.

\end{abstract}

\noindent \textbf{Keywords:} Acyclic edge coloring, acyclic edge chromatic number, planar graphs.

\section{Introduction}

All graphs considered in this paper are finite and simple. A proper \emph{edge coloring} of $G=(V,E)$ is a map $c: E\rightarrow C$ (where $C$ is the set of available $colors$ ) with $c(e) \neq c(f)$ for any adjacent edges $e$,$f$. The minimum number of colors needed to properly color the edges of $G$, is called the chromatic index of $G$ and is denoted by $\chi'(G)$. A proper edge coloring c is called acyclic if there are no bichromatic cycles in the graph. In other words an edge coloring is acyclic if the union of any two color classes induces a set of paths (i.e., linear forest) in $G$. The \emph{acyclic edge chromatic number} (also called \emph{acyclic chromatic index}), denoted by $a'(G)$, is the minimum number of colors required to acyclically edge color $G$. The concept of \emph{acyclic coloring} of a graph was introduced by Gr\"unbaum \cite{Grun}. The \emph{acyclic chromatic index} and its vertex analogue can be used to bound other parameters like \emph{oriented chromatic number} and \emph{star chromatic number} of a graph, both of which have many practical applications, for example, in wavelength routing in optical networks ( \cite{ART}, \cite{KSZ} ). Let $\Delta=\Delta(G)$ denote the maximum degree of a vertex in graph $G$. By Vizing's theorem, we have $\Delta \le \chi'(G) \le \Delta +1 $(see \cite{Diest} for proof). Since any acyclic edge coloring is also proper, we have $a'(G)\ge\chi'(G)\ge\Delta$. \newline

It has been conjectured by Alon, Sudakov and Zaks \cite{ASZ} (and much earlier by Fiamcik \cite{Fiam}) that $a'(G)\le\Delta+2$ for any $G$. Using probabilistic arguments Alon, McDiarmid and Reed \cite{AMR} proved that $a'(G)\le60\Delta$. The best known result up to now for arbitrary graph, is by Molloy and Reed  \cite{MolReed} who showed that $a'(G)\le16\Delta$. Muthu, Narayanan and Subramanian \cite{MNS1} proved that $a'(G)\le4.52\Delta$ for graphs $G$ of girth at least 220 (\emph{Girth} is the length of a shortest cycle in a graph).\newline

Though the best known upper bound for general case is far from the conjectured $\Delta+2$, the conjecture has been shown to be true for some special classes of graphs. Alon, Sudakov and Zaks \cite{ASZ} proved that there exists a constant $k$ such that $a'(G)\le\Delta+2$ for any graph $G$ whose girth is at least $k\Delta\log\Delta$. They also proved that $a'(G)\le\Delta+2$ for almost all $\Delta$-regular graphs. This result was improved by Ne\v set\v ril and Wormald \cite{NesWorm} who showed that for a random $\Delta$-regular graph $a'(G)\le \Delta+1$. Muthu, Narayanan and Subramanian proved the conjecture for grid-like graphs \cite{MNS2}. In fact they gave a better bound of $\Delta+1$ for these class of graphs. From Burnstein's \cite{Burn} result it follows that the conjecture is true for subcubic graphs. Skulrattankulchai \cite{Skul} gave a polynomial time algorithm to color a subcubic graph using $\Delta+2 = 5$ colors. Fiamcik \cite{Fiam1}, \cite{Fiam2} proved that every subcubic graph, except for $K_4$ and $K_{3,3}$, is acyclically edge colorable using $4$ colors.

Determining $a'(G)$ is a hard problem both from theoretical and algorithmic points of view. Even for the simple and highly structured class of complete graphs, the value of $a'(G)$ is still not determined exactly. It has also been shown by Alon and Zaks \cite{AZ} that determining whether $a'(G)\le3$ is NP-complete for an arbitrary graph $G$. The vertex version of this problem has also been extensively studied ( see \cite{Grun}, \cite{Burn}, \cite{Boro}). A generalization of the acyclic edge chromatic number has been studied: The \emph{r-acyclic edge chromatic number} $a'_r(G)$ is the minimum number of colors required to color the edges of the graph $G$ such that every cycle $C$ of $G$ has at least min\{$\vert C \vert$,$r$\} colors ( see \cite{GeRa}, \cite{GrePi}).

~~~~~~~~~

\noindent\textbf{Our Result:} The acyclic chromatic index of planar graphs has been studied previously. Fiedorowicz, Hauszczak and Narayanan \cite{AMN} gave an upper bound of $2\Delta+29$ for planar graphs. Independently Hou, Wu, GuiZhen Liu and Bin Liu \cite{JWu1} gave an upper bound of $max(2\Delta - 2,\Delta+22)$, which is the best known result up to now for planar graphs. Note that for $\Delta \ge 24$, it is equal to $2\Delta-2$. In this paper, we prove the following theorem,

\begin{thm}
\label{thm:thm1}
If $G$ is a planar graph, then $a'(G)\le \Delta +12$.
\end{thm}

The acyclic chromatic index of special classes of planar graphs based on girth and absence of short cycles as also been studied. In \cite{AMN}, an upper bound of $\Delta+6$ for triangle free planar graphs has been proved. In \cite{JWu1} an upper bound of $\Delta+2$ for planar graphs of girth at least five has been proved. Fiedorowicz and Borowiecki \cite{AFMB} proved an upper bound of $\Delta+1$ for planar graphs of girth at least six and an upper bound of $\Delta+15$ for planar graphs without four cycles.

Our proof is constructive and yields an efficient polynomial time algorithm. We have presented the proof in a non-algorithmic way. But it is easy to extract the underlying algorithm from it.

\section{Preliminaries}

Let $G=(V,E)$ be a simple, finite and connected graph of $n$ vertices and $m$ edges. Let $x \in V$. Then $N_{G}(x)$ will denote the neighbours of $x$ in $G$. For an edge $e \in E$, $G-e$ will denote the graph obtained by deletion of the edge $e$. For $x,y \in V$, when $e=(x,y)=xy$, we may use $G-\{xy\}$ instead of $G-e$. Let $c:E\rightarrow \{1,2,\ldots,k\}$ be an \emph{acyclic edge coloring} of $G$. For an edge $e\in E$, $c(e)$ will denote the color given to $e$ with respect to the coloring $c$. For $x,y \in V$, when $e=(x,y)=xy$ we may use $c(x,y)$ instead of $c(e)$. For $S \subseteq V$, we denote the induced subgraph on $S$ by $G[S]$.

Many of the definitions, facts and lemmas that we develop in this section are already present in our earlier papers \cite{MBSC2}, \cite{MBSC4}. We include them here for the sake of completeness. The proofs of the lemmas will be omitted whenever it is available in \cite{MBSC2}, \cite{MBSC4}.

~~~~~~

\noindent \textbf{Partial Coloring:} Let H be a subgraph of $G$. Then an edge coloring $c'$ of $H$ is also a partial coloring of $G$. Note that $H$ can be $G$ itself. Thus a coloring $c$ of $G$ itself can be considered a partial coloring. A partial coloring $c$ of $G$ is said to be a proper partial coloring if $c$ is proper. A proper partial coloring $c$ is called acyclic if there are no bichromatic cycles in the graph. Sometimes we also use the word valid coloring instead of acyclic coloring. Note that with respect to a partial coloring $c$, $c(e)$ may not be defined for an edge $e$. So, whenever we use $c(e)$, we are considering an edge $e$ for which $c(e)$ is defined, though we may not always explicitly mention it.

Let $c$ be a partial coloring of $G$. We denote the set of colors in the partial coloring $c$ by $C = \{1,2,\ldots,k\}$. For any vertex $u \in V(G)$, we define $F_u(c) =\{c(u,z) \vert z \in N_{G}(u)\}$. For an edge $ab \in E$, we define $S_{ab}(c) = F_b - \{c(a,b)\}$. Note that $S_{ab}(c)$ need not be the same as $S_{ba}(c)$. We will abbreviate the notation to $F_u$ and $S_{ab}$ when the coloring $c$ is understood from the context.

To prove the main result, we plan to use contradiction. Let $G$ be the minimum counter example with respect to the number of edges for the statement in the theorems that we plan to prove.  Let $G =(V,E)$ be a graph on $m$ edges where $m \ge 1$. We will remove an edge $e=(x,y)$ from $G$ and get a graph $G'=(V,E')$. By the minimality of $G$, the graph $G'$ will have an acyclic edge coloring $c:E'\rightarrow \{1,2,\ldots,t\}$, where $t$ is the claimed upper bound for $a'(G)$. Our intention will be to extend the coloring $c$ of $G'$ to $G$ by assigning an appropriate color for the edge $e$ thereby contradicting the assumption that $G$ is a minimum counter example.

~~~~~

The following definitions arise out of our attempt to understand what may prevent us from extending a partial coloring of $G-e$ to $G$.

\noindent \textbf{Maximal bichromatic Path:} An ($\alpha$,$\beta$)-maximal bichromatic path with respect to a partial coloring $c$ of $G$ is a maximal path consisting of edges that are colored using the colors $\alpha$ and $\beta$ alternatingly. An ($\alpha$,$\beta$,$a$,$b$)-maximal bichromatic path is an ($\alpha$,$\beta$)-maximal bichromatic path which starts at the vertex $a$ with an edge colored $\alpha$ and ends at $b$. We emphasize that the edge of the ($\alpha$,$\beta$,$a$,$b$)-maximal bichromatic path incident on vertex $a$ is colored $\alpha$ and the edge incident on vertex $b$ can be colored either $\alpha$ or $\beta$. Thus the notations ($\alpha$,$\beta$,$a$,$b$) and ($\alpha$,$\beta$,$b$,$a$) have different meanings. Also note that any maximal bichromatic path will have at least two edges. The following fact is obvious from the definition of proper edge coloring:

\begin{fact}
\label{fact:fact1}
Given a pair of colors $\alpha$ and $\beta$ of a proper coloring $c$ of $G$, there can be at most one maximal ($\alpha$,$\beta$)-bichromatic path containing a particular vertex $v$, with respect to $c$.
\end{fact}

A color $\alpha \neq c(e)$ is a \emph{candidate} for an edge \emph{e} in $G$ with respect to a partial coloring $c$ of $G$ if none of the adjacent edges of \emph{e} are colored $\alpha$. A candidate color $\alpha$ is \emph{valid} for an edge \emph{e} if assigning the color $\alpha$ to \emph{e} does not result in any bichromatic cycle in $G$.

Let $e=(a,b)$ be an edge in $G$. Note that any color $\beta \notin F_a \cup F_b$ is a candidate color for the edge $ab$ in $G$ with respect to the partial coloring $c$ of $G$. A sufficient condition for a candidate color being valid is captured in the Lemma below (See Appendix for proof):

\begin{lem}
\label{lem:lem1}
\cite{MBSC2}
A candidate color for an edge $e=ab$, is valid if $(F_a \cap F_b) - \{c(a,b)\} = (S_{ab} \cap S_{ba}) = \emptyset$.
\end{lem}

Now even if $S_{ab} \cap S_{ba} \neq \emptyset$, a candidate color $\beta$ may be valid. But if $\beta$ is not valid, then what may be the reason? It is clear that color $\beta$ is not $valid$ if and only if there exists $\alpha \neq \beta$ such that a ($\alpha$,$\beta$)-bichromatic cycle gets formed if we assign color $\beta$ to the edge $e$. In other words, if and only if, with respect to coloring $c$ of $G$ there existed a ($\alpha$,$\beta$,$a$,$b$) maximal bichromatic path with $\alpha$ being the color given to the first and last edge of this path. Such paths play an important role in our proofs. We call them $critical\ paths$. It is formally defined below: \newline

\noindent\textbf{Critical Path:} Let $ab \in E$ and $c$ be a partial coloring of $G$. Then a $(\alpha,\beta,$a$,$b$)$ maximal bichromatic path which starts out from the vertex $a$ via an edge colored  $\alpha$ and ends at the vertex $b$ via an edge colored $\alpha$ is called an $(\alpha,\beta,ab)$ critical path. Note that any critical path will be of odd length. Moreover the smallest length possible is three.

~~~~~~~~~
%

An obvious strategy to extend a valid partial coloring $c$ of $G$ would be to try to assign one of the candidate colors to an uncolored edge $e$. The condition that a candidate color being not valid for the edge $e$ is captured in the following fact.

\begin{fact}
\label{fact:fact2}
Let $c$ be a partial coloring of $G$. A candidate color $\beta$ is not $valid$ for the edge $e=(a,b)$ if and only if $\exists \alpha \in S_{ab} \cap S_{ba}$ such that there is a $(\alpha,\beta,ab)$  critical path in $G$ with respect to the coloring $c$.
\end{fact}

%

\noindent \textbf{Color Exchange:} Let $c$ be a partial coloring of $G$. Let $u,i,j \in V(G)$ and $ui,uj \in E(G)$. We define $Color\ Exchange$ with respect to the edge $ui$ and $uj$, as the modification of the current partial coloring $c$ by exchanging the colors of the edges $ui$ and $uj$ to get a partial coloring $c'$, i.e., $c'(u,i)=c(u,j)$, $c'(u,j)=c(u,i)$ and $c'(e)=c(e)$ for all other edges $e$ in $G$. The color exchange with respect to the edges $ui$ and $uj$ is said to be proper if the coloring obtained after the exchange is proper. The color exchange with respect to the edges $ui$ and $uj$ is $valid$ if and only if the coloring obtained after the exchange is acyclic. The following fact is obvious:

\begin{fact}
\label{fact:fact3}
Let $c'$ be the partial coloring obtained from a valid partial coloring $c$ by the color exchange with respect to the edges $ui$ and $uj$. Then the partial coloring $c'$ will be proper if and only if $c(u,i) \notin S_{uj}$ and $c(u,j) \notin S_{ui}$.
\end{fact}

~~~~~~

\noindent The color exchange is useful in breaking some critical paths as is clear from the following lemma (See Appendix for proof):

\begin{lem}
\label{lem:lem2}
\cite{MBSC2}, \cite{MBSC4}
Let $u,i,j,a,b \in V(G)$, $ui,uj,ab \in E$. Also let $\{\lambda,\xi\} \in C$ such that $\{\lambda,\xi\} \cap \{c(u,i),c(u,j)\} \neq \emptyset$ and $\{i,j\} \cap \{a,b\} = \emptyset$. Suppose there exists an ($\lambda$,$\xi$,$ab$)-critical path that contains vertex $u$, with respect to a valid partial coloring $c$ of $G$. Let $c'$ be the partial coloring obtained from $c$ by the color exchange with respect to the edges $ui$ and $uj$. If $c'$ is proper, then there will not be any ($\lambda$,$\xi$,$ab$)-critical path in $G$ with respect to the partial coloring $c'$.
\end{lem}

%

\noindent \textbf{Multisets and Multiset Operations:}
Recall that a multiset is a \emph{generalized} set where a member can appear multiple times in the set. If an element $x$ appears $t$ times in the multiset $S$, then we say that multiplicity of $x$ in $S$ is $t$. In notation $mult_{S}(x)=t$. The cardinality of a finite multiset $S$, denoted by $\parallel S \parallel$ is defined as $\parallel S \parallel=\sum_{x \in S}mult_{S}(x)$.Let $S_1$ and $S_2$ be two multisets. The reader may note that there are various possible ways to define union of $S_1$ and $S_2$. For the purpose of this paper we will define one such union notion- which we call as the $join$ of $S_1$ and $S_2$, denoted as $S_1 \uplus S_2$. The multiset $S_1 \uplus S_2$ will have all the members of $S_1$ as well as $S_2$. For a member $x \in S_1 \uplus S_2$, $mult_{S_1 \uplus S_2}(x) = mult_{S_1}(x) + mult_{S_2}(x)$. Clearly $\parallel S_1 \uplus S_2 \parallel= \parallel S_1 \parallel + \parallel S_2 \parallel$. We also need a specially defined notion of the multiset difference of $S_1$ and $S_2$, denoted by $S_1 \setminus S_2$. It is the multiset of elements of $S_1$ which are not in $S_2$, i.e., $x \in S_1 \setminus S_2$ iff $x \in S_1$ but $x \notin S_2$ and $mult_{S_1 \setminus S_2}(x)= mult_{S_1}(x)$.

\section{Proof of Theorem 1}

\begin{proof}

A well-known strategy that is used in proving coloring theorems in the context of planar graphs is to make use of induction combined with the fact that there are some \emph{unavoidable} configurations in any planar graph. Typically the existence of these \emph{unavoidable} configurations are proved using the so called \emph{charging and discharging argument} (See \cite{Salavatipour}, for a comprehensive exposition). Loosely speaking, for the purpose of this paper, a \emph{configuration} is a set $\{v\} \cup N(v)$, where $v$ is some vertex in $G$, along with some information regarding the degrees of the vertices in $\{v\} \cup N(v)$. For example, the following lemma illustrates how certain unavoidable configurations appear in a planar graph.

\begin{lem}
\label{lem:lem4}
\cite{HeuvelMcG}
Let G be a simple planar graph with $\delta \ge 2$, where $\delta$ is the minimum degree of graph $G$. Then there exists a vertex $v$ in $G$ with exactly $d(v)=k$ neighbours $v_1,v_2,\dots,v_k$ with $d(v_1)\le d(v_2) \le \ldots \le d(v_k)$ such that at least one of the following is true:
\begin{enumerate}
\item[(A1)] $k=2$,
\item[(A2)] $k=3$ and $d(v_1) \le 11$,
\item[(A3)] $k=4$ and $d(v_1) \le 7$, $d(v_2) \le 11$,
\item[(A4)] $k=5$ and $d(v_1) \le 6$, $d(v_2) \le 7$, $d(v_3) \le 11$.
\end{enumerate}
\end{lem}

~~~~~~

Let graph $G$ be a minimum counter example with respect to the number of edges for the statement in Theorem \ref{thm:thm1}. From $Lemma$ \ref{lem:lem4} we know that there exists a vertex $v$ in $G$ such that it belongs to one of the configurations $A1$-$A4$. We now delete the edge $vv_1$ to get a graph $G'$, where $v$ and $v_1$ are as in $Lemma$ \ref{lem:lem4}. Since $G$ was the minimum counter example, $G'$ has an acyclic edge coloring using $\Delta(G')+12$ colors. Let $c'$ be such a coloring. Now if $\Delta(G') < \Delta(G)$, then we have at least one extra color for $G$ and we can assign that color to edge $vv_1$ to get a valid coloring of $G$, a contradiction to the fact that $G$ is a counter example. Thus we have $\Delta(G')=\Delta(G)=\Delta$. To prove the theorem for $G$, we may assume that $G$ is $2$-$connected$ since if there are cut vertices in $G$, the acyclic edge coloring of the blocks $B_1,B_2\ldots B_k$ of $G$ can easily be extended to $G$. Thus we have, $\delta(G)\ge 2$. We present the proof in two parts based on which configuration the vertex $v$ belongs to - The first part deals with the case when there exists a vertex $v$ that belongs to configuration $A2$,$A3$ or $A4$ and the second part deals with the case when there does not exists any vertex $v$ in $G$ that belongs to configuration $A2$,$A3$ or $A4$.

~~~~

\subsection{There exists a vertex $v$ that belongs to configuration $A2$,$A3$ or $A4$}

\begin{clm}
\label{clm:clm11}
For any valid coloring $c'$ of $G'$, $\vert F_v \cap F_{v_1} \vert \ge 2$.
\end{clm}
\begin{proof}
Suppose not. The case $\vert F_v \cap F_{v_1} \vert= 0$ is trivial. The reader can verify from close examination of configurations $A2$-$A4$ that $\vert F_v \cup F_{v_1} \vert$ will be maximum for configuration $A2$ and therefore $\vert F_v \cup F_{v_1} \vert = \vert F_v \vert + \vert F_{v_1} \vert \le 2 + 10 = 12$. Thus there are $\Delta$ candidate colors for the edge $vv_1$ and by $Lemma$ \ref{lem:lem1} all the candidate colors are valid, a contradiction to the assumption that $G$ is a counter example. Thus we have $\vert F_v \cap F_{v_1} \vert= 1$. In this case it is easy to see that $\vert F_v \cup F_{v_1} \vert = \vert F_v \vert +\vert F_{v_1} \vert- \vert F_v \cap F_{v_1} \vert \le 11$ and hence there are at least $\Delta + 1$ candidate colors for the edge $vv_1$. Let $F_v \cap F_{v_1} = \{\alpha\}$ and let $u \in N(v)$ be a vertex such that $c'(v,u)=\alpha$. Now if none of the $\Delta +1$ candidate colors are valid for the edge $vv_1$, then by $Fact$ \ref{fact:fact2}, for each $\gamma \in C-(F_v \cup F_{v_1})$, there exists a $(\alpha,\gamma,vv_1)$ critical path. Since $c'(v,u)=\alpha$, we have all the critical paths passing through the vertex $u$ and hence $S_{vu} \subseteq C-(F_v \cup F_{v_1})$. This implies that $\vert S_{vu} \vert \ge \vert C-(F_v \cup F_{v_1}) \vert \ge (\Delta+12)-11 = \Delta+1$, a contradiction since $\vert S_{vu} \vert \le \Delta-1$. Thus we have a valid color for the edge $vv_1$, a contradiction to the assumption that $G$ is a counter example. Thus $\vert F_v \cap F_{v_1} \vert \ge 2$.
\end{proof}

Let $S_v$ be a multiset defined as $S_v= S_{vv_2} \uplus S_{vv_3} \uplus \ldots \uplus S_{vv_k}$. In view of $Claim$ \ref{clm:clm11} and Lemma \ref{lem:lem4}, $2 \le \vert F_v \cap F_{v_1} \vert \le 4$. We consider each case separately.

\subsection*{case 1: $\vert F_v \cap F_{v_1} \vert= 2$}

Let $F_v \cap F_{v_1}=\{1,2\}$ and let $v',v'' \in N_{G'}(v)$ and $u',u'' \in N_{G'}(v_1)$ be such that $c'(v,v')=c'(v_1,u')=1$ and $c'(v,v'')= c'(v_1,u'')=2$. It is easy to see that $\vert F_v \cup F_{v_1} \vert \le 10 $. Thus there are at least $\Delta+2$ candidate colors for the edge $vv_1$. If any of the candidate colors are valid for the edge $vv_1$, we are done. Thus none of the candidate colors are valid for the edge $vv_1$. This implies that there exist either a $(1,\theta,vv_1)$ or $(2,\theta,vv_1)$ critical path for each candidate color $\theta$.

\begin{clm}
\label{clm:clm12}
With respect to the coloring $c'$, the multiset $S_v$ contains at least $\vert F_{v_1} \vert - 1$ colors from $F_{v_1}$.
\end{clm}
\begin{proof}
Suppose not. Then there are at least two colors in $F_{v_1}$ which are not in $S_v$. Let $\nu$ and $\mu$ be any two such colors. Now assign colors $\nu$ and $\mu$ to the edges $vv'$ and $vv''$ respectively to get a coloring $c''$. Now since $\nu, \mu \notin S_v$, we have $\nu \notin S_{vv'}$ and $\mu \notin S_{vv''}$. Moreover $\mu,\nu \notin \{1,2\}$. Thus the recoloring $c''$ is proper. Now we claim that the coloring $c''$ is acyclic also. Suppose not. Then there has to be a bichromatic cycle containing at least one of the colors $\nu$ and $\mu$. Clearly this cannot be a $(\nu,\mu)$ bichromatic cycle since $\mu \notin S_{vv'}$. Therefore it has to be a $(\nu,\lambda)$ or $(\mu,\lambda)$ bichromatic cycle where $\lambda \in F_v(c'')-\{\nu,\mu\}$. Let $u$ be a vertex such that $c''(v,u)=\lambda$. This means that there was already a $(\lambda,\nu,vv')$ or $(\lambda,\mu,vv'')$ critical path with respect to the coloring $c'$. This implies that $\nu \in S_{vu}$ or $\mu \in S_{vu}$, implying that $\nu \in S_{v}$ or $\mu \in S_{v}$, a contradiction. Thus the coloring $c''$ is acyclic. Let $u_1,u_2 \in N_{G'}(v_1)$ be such that $c''(v_1,u_1)=\nu$ and $c''(v_1,u_2)=\mu$.

Note that $\vert F_{v} \cup F_{v_1} \vert \le 10$ (The maximum value of $\vert F_{v} \cup F_{v_1} \vert$ is attained when the graph has configuration $A2$). Therefore there are at least $\Delta+2$ candidate colors for the edge $vv_1$. If any of the candidate colors are valid for the edge $vv_1$, then we are done as this is a contradiction to the assumption that $G$ is a counter example. Thus none of the candidate colors are valid for the edge $vv_1$ and therefore there exist either a $(\nu,\theta,vv_1)$ or $(\mu,\theta,vv_1)$ critical path for each candidate color $\theta$. Let $C_{\nu}$ and $C_{\mu}$ respectively be the set of candidate colors which are forming critical paths with colors $\nu$ and $\mu$. Then clearly $C_{\nu} \subseteq S_{v_1u_1}$ and $C_{\mu} \subseteq S_{v_1u_2}$ since $c''(v_1,u_1)=\nu$ and $c''(v_1,u_2)=\mu$. Now we \emph{exchange the colors} of the edges $vv'$ and $vv''$ to get a modified coloring $c$. Note that $c$ is proper since $\mu \notin S_{vv'}$ and $\nu \notin S_{vv''}$. By Lemma \ref{lem:lem2}, all $(\nu,\beta,vv_1)$ critical paths where $\beta \in C_{\nu}$ and all $(\mu,\gamma,vv_1)$ critical paths where $\gamma \in C_{\mu}$ are broken. Now if none of the colors in $C_{\nu}$ are valid for edge $vv_1$, then it means that for each $\beta \in C_{\nu}$, there exists a $(\mu,\beta,vv_1)$ critical path with respect to coloring $c$, implying that $C_{\nu} \subseteq S_{v_1u_2}$. Since the recoloring involved no candidate colors, we still have $C_{\mu} \subseteq S_{v_1u_2}$. Thus we have  $(C_{\nu} \cup C_{\mu}) \subseteq S_{v_1u_2}$. But $\vert C_{\nu} \cup C_{\mu} \vert \ge \Delta +2$ which implies that $\vert S_{v_1u_2} \vert \ge \Delta +2$, a contradiction since $\vert S_{v_1u_2} \vert \le \Delta -1$.
\end{proof}

\begin{clm}
\label{clm:clm13}
With respect to the coloring $c'$, there exists at least two colors $\alpha$ and $\beta$ in $C-F_{v_1}$ with multiplicity at most one in $S_v$.
\end{clm}
\begin{proof}
In view of Claim \ref{clm:clm12} we have $\sum_{x \in C-F_v} mult_{S_v}(x) = \parallel S_v \parallel - (\vert F_v \vert -1)$. Thus if $\parallel S_v \parallel - (\vert F_{v_1} \vert -1) \le 2\vert (C-F_{v_1}) \vert -3$, then there exist at least two colors $\alpha$ and $\beta$ in $C-F_{v_1}$ with multiplicity at most one in $S_v$. Thus it is enough to prove $\parallel S_v \parallel \le 2\vert C \vert - \vert F_{v_1} \vert -4 \le 2\Delta + 24 -\vert F_{v_1} -4 = 2\Delta + 20 -\vert F_{v_1} \vert$. Now we can easily verify that $\parallel S_v \parallel + \vert F_{v_1} \vert \le 2 \Delta +20$ for configurations $A2-A4$ as follows:

\begin{itemize}
\item For $A2$, $\parallel S_v \parallel + \vert F_{v_1} \vert \le  (d(v_2)-1) + (d(v_3)-1) + \vert F_{v_1} \vert = (\Delta -1) + (\Delta -1) \ \ + 10 = 2\Delta +8$.

\item For $A3$, $\parallel S_v \parallel + \vert F_{v_1} \vert \le  (d(v_2)-1) + (d(v_3)-1) + (d(v_4)-1) + \vert F_{v_1} \vert = 10+ (\Delta -1) + (\Delta -1) \ \ + 6 = 2\Delta +14$.

\item For $A4$, $\parallel S_v \parallel + \vert F_{v_1} \vert \le  (d(v_2)-1) + (d(v_3)-1) + (d(v_4)-1) + (d(v_5)-1)+ \vert F_{v_1} \vert = 6 + 10 +(\Delta -1) + (\Delta -1)\ \ +\ \ 5  = 2\Delta +19$.
\end{itemize}

\end{proof}

\noindent The colors $\alpha$ and $\beta$ of $Claim$ \ref{clm:clm13} are crucial to the proof. Now we make another claim regarding $\alpha$ and $\beta$:

\begin{clm}
\label{clm:clm14}
With respect to the coloring $c'$, $\alpha$ and $\beta \in F_v$.
\end{clm}
\begin{proof}
Without loss of generality, let $\alpha \notin F_v$. Then recalling that $\alpha \notin F_{v_1}$,  $\alpha$ is a candidate for the edge $vv_1$. If it is not valid, then there exists either a $(1,\alpha,vv_1)$ or $(2,\alpha,vv_1)$ critical path with respect to $c'$. Since the multiplicity of $\alpha$ in $S_{v}$ is at most one, we have the color $\alpha$ in exactly one of $S_{vv'}$ or $S_{vv''}$. Without loss of generality let $\alpha \in S_{vv''}$. Hence there exists either a $(2,\alpha,vv_1)$ critical path with respect to $c'$.

Now recolor the edge $vv'$ with color $\alpha$ to get a coloring $c$. It is obvious that the recoloring $c$ is proper since $\alpha \notin F_v(c')$ and $\alpha \notin S_{vv'}(c')$. It is also valid since if a bichromatic cycle gets formed due to this recoloring, it has to be a $(\alpha,\gamma)$ bichromatic cycle for some $\gamma \in F_v(c) - c(v,v')$. Let $a \in N_{G'}(v)$ be such that $c(v,a)=\gamma$. Then the $(\alpha,\gamma)$ bichromatic cycle should contain the edge $va$ and therefore $\gamma \in S_{va}$ with respect to $c$. But we know that $v''$ is the only vertex in $N_{G'}(v)$ such that $\alpha \in S_{vv''}$. Therefore $a=v''$. This implies that $\gamma =2$ and there existed a $(2,\alpha,vv')$ critical path with respect to the coloring $c'$. This is a contradiction to the Fact \ref{fact:fact1} since there already existed a $(2,\alpha,vv_1)$ critical path with respect to the coloring $c'$. Thus the recoloring $c$ is valid. Now with respect to the coloring $c$, $\vert F_v \cap F_{v_1} \vert= 1$, a contradiction to $Claim$ \ref{clm:clm11}.
\end{proof}

Note that $\alpha, \beta \notin \{1,2\}$ since $\alpha, \beta \notin F_{v_1}$. In view of $Claim$ \ref{clm:clm14}, we have $\{1,2,\alpha,\beta\} \subseteq F_v$ and thus  $\vert F_v \vert \ge 4$, which implies that $d(v) \ge 5$. Thus the vertex $v$ belongs to configuration $A4$. Therefore $d(v)=5$ and $F_v=\{1,2,\alpha,\beta\}$. There are at least $\Delta+12 - (5+4-2)=\Delta+5$ candidate colors for the edge $vv_1$. Also recall that $d(v_2) \le 7$, $c'(v,v')=c'(v_1,u')=1$ and $c'(v,v'')= c'(v_1,u'')=2$.

\begin{clm}
\label{clm:clm15}
With respect to the coloring $c'$, $v_2 \notin \{v',v''\}$.
\end{clm}
\begin{proof}
Suppose not. Then without loss of generality let $v_2 = v'$ and $c'(v,v_2)=1$. Now if none of the $\Delta+5$ candidate colors are valid for the edge $vv_1$, then they all form critical paths that contain either the edge $vv'$ or the edge $vv''$. Now $\vert S_{vv'} \vert$ + $\vert S_{vv''} \vert \le 6 +\Delta -1 = \Delta+5$. Since each of the $\Delta+5$ candidate colors has to be present in either in $S_{vv'}$ or $S_{vv''}$, we infer that $S_{vv''} \cup S_{vv'}$ is exactly the set of candidate colors, i.e., $\vert S_{vv'} \vert$ + $\vert S_{vv''} \vert$ = $\Delta+5$. This requires that $\vert S_{vv'} \vert =6$, $\vert S_{vv''} \vert = \Delta-1$ and $S_{vv''} \cap S_{vv'} = \emptyset$. Since for each $\gamma \in S_{vv''}$, we have $(2,\gamma,vv_1)$ critical path containing $u''$, we can infer that $S_{vv''} \subseteq S_{v_1u''}$ (Recall that $c'(v_1,u'')=2$). But since $\vert S_{v_1u''} \vert \le \Delta-1$, we infer $S_{vv''} = S_{v_1u''}$. Thus we have $S_{v_1u''} \cap S_{vv'} = S_{vv''} \cap S_{vv'} = \emptyset$.

Now we exchange the colors of the edges $vv'$ and $vv''$ to get a coloring $c$ i.e., $c(v,v')=2$ and $c(v,v'')=1$. The coloring $c$ is proper since $2 \notin S_{vv'}(c')$ and $1 \notin S_{vv''}(c')$ (Recall that $S_{vv'}(c')$ and $S_{vv''}(c')$ contain only candidate colors). The coloring is also valid: If a bichromatic cycle gets formed it has to be a $(1,\eta)$ or $(2,\eta)$ bichromatic cycle where $\eta \in F_v$. Clearly it cannot be a $(1,2)$ bichromatic cycle since $1 \notin S_{vv'}(c)$ and therefore $\eta= \alpha$ or $\beta$ (Recall that $F_v=\{1,2,\alpha,\beta\}$). This implies that either $\alpha$ or $\beta$ belongs to $S_{vv'} \cup S_{vv''}$. But we know that $S_{vv'} \cup S_{vv''}$ is exactly the set of candidate colors for the edge $vv_1$, a contradiction since $\alpha,\beta \in F_v$ cannot be candidate colors for the edge $vv_1$. Therefore the coloring $c$ is acyclic. By $Lemma$ \ref{lem:lem2}, all the existing critical paths are broken. Now consider a color $\gamma \in S_{vv'}$. If it is still not valid then there has to be a $(2,\gamma,vv_1)$ critical path since $c(v,v')=2$ and $\gamma \notin S_{vv''}(c)$. This implies that $\gamma \in S_{v_1u''}(c)$, a contradiction since $S_{v_1u''}(c) \cap S_{vv'}(c) = \emptyset$. Thus we have a valid color for the edge $vv_1$, a contradiction to the assumption that $G$ is a counter example. Thus $v_2 \notin \{v',v''\}$.
\end{proof}

From $Claim$ \ref{clm:clm15}, we infer that $c'(v,v_2) \notin F_v \cap F_{v_1}$ since $F_v \cap F_{v_1}=\{c'(v,v'),c(v,v'')\}=\{1,2\}$. Therefore we have $c(v,v_2) \in \{\alpha,\beta\}$ since $F_v = \{1,2,\alpha,\beta\}$. Without loss of generality let $c(v,v_2)=\alpha$. We know that the color $\beta$ can be in at most one of $S_{vv'}$ and $S_{vv''}$ by Claim \ref{clm:clm13}. Now let $v'$ be such that $\beta \notin S_{vv'}$. Note that $C-(S_{vv'} \cup F_v \cup F_{v_1})\neq \emptyset$ since $\vert S_{vv'} \cup F_v \cup F_{v_1} \vert \le \Delta-1 +4+5-2 = \Delta+6$. Assign a color $\theta \in C-(S_{vv'} \cup F_v \cup F_{v_1})$ to the edge $vv'$ to get a coloring $c''$. If it is valid, then let $c=c''$.

If the recoloring is not valid then there has to be a bichromatic cycle created due to the recoloring. Now the bichromatic cycle should involve one of the colors $2$, $\alpha$, $\beta$ along with $\theta$. Since the bichromatic cycle contains a color from $S_{vv'}$ and $\beta \notin S_{vv'}$, it cannot be a $(\theta,\beta)$ bichromatic cycle. Now with respect to the coloring $c'$, color $\theta$ was not valid for the edge $vv_1$ implying that there existed either a  $(1,\theta,vv_1)$ or a $(2,\theta,vv_1)$ critical path. But $(1,\theta,vv_1)$ critical path was not possible since $\theta \notin S_{vv'}$ by the choice of $\theta$. Thus there existed a $(2,\theta,vv_1)$ critical path with respect to $c'$. Thus by Fact $\ref{fact:fact1}$, there cannot be a $(2, \theta,vv')$ critical path with respect to $c'$ and hence there cannot be a $(2,\theta)$ bichromatic cycle in $c''$ formed due to the recoloring. Thus if there is a bichromatic cycle formed, then it has to be a $(\alpha,\theta)$ bichromatic cycle,  which implies that $\alpha \in S_{vv'}$.

Now taking into account the fact that $\alpha$ is in $S_{vv'}$ as well as $F_v$, we get $\vert S_{vv'} \cup F_v \cup F_{v_1} \vert \le \Delta-1 +4+5-2-1 = \Delta+5$ and therefore  $\vert S_{vv'} \cup F_v \cup F_{v_1} \cup S_{vv_2} \vert \le \Delta + 5 +6 = \Delta +11$. Thus $C-(S_{vv'} \cup F_v \cup F_{v_1} \cup S_{vv_2}) \neq \emptyset$. Now recolor the edge $vv'$ using a color $\gamma \in C-(S_{vv'} \cup F_v \cup F_{v_1} \cup S_{vv_2})$ to get a coloring $c$. Clearly the recoloring is proper. It is also valid since if a bichromatic cycle gets formed it has to be a $(\alpha,\gamma)$ bichromatic cycle (Note that the $(2,\gamma)$ and $(\beta,\gamma)$ bichromatic cycles are argued out as before). But $\gamma \notin S_{vv_2}$, a contradiction. Thus the coloring $c$ is acyclic.

With respect to the coloring $c$ we have $\vert F_v \cap F_{v_1} \vert= 1$, a contradiction to Claim \ref{clm:clm11}.

~~~~~


\subsection*{case 2: $\vert F_v \cap F_{v_1} \vert = 3$}

Note that in this case $\vert F_v \vert \ge 3$ and therefore $d(v) \ge 4$. Thus $v$ belongs to either configuration $A3$ or $A4$. Let $S'_v$ be a multiset defined by $S'_v = S_v \setminus (F_{v_1} \cup F_{v})$. Let $v',v'', v''' \in N_{G'}(v)$ be such that $\{c(v,v'),c(v,v''),c(v,v''')\} = F_v \cap F_{v_1}$. Also let $c(v,v')=1$, $c(v,v'')=2$ and $c(v,v''')=3$.

\begin{clm}
\label{clm:clm16}
With respect to $c'$, $\parallel S'_v \parallel \le 2\Delta +11$.
\end{clm}
\begin{proof}
When $d(v)=4$, it is clear that $\parallel S'_v \parallel \le (d(v_2)-1) + (d(v_3)-1) + (d(v_4)-1) \le 10 + \Delta-1 + \Delta-1 = 2\Delta +8$. On the other hand when $d(v)=5$, try to recolor one of the edges $vv'$, $vv''$, $vv'''$ using a color in $C-(F_v \cup F_{v_1})$. There are $\Delta +6$ colors in $C-(F_v \cup F_{v_1})$ and if any of these colors is valid for one of $vv'$, $vv''$ or $vv'''$, then the situation reduces to case 1 i.e., $\vert F_v \cap F_{v_1} \vert = 2$. Otherwise there has to be a bichromatic cycle formed during each recoloring. Since such a bichromatic cycle has to be $(\gamma_1,\gamma_2)$ bichromatic cycle where $\gamma_1$ is the color used in the recoloring and $\gamma_2 \in F_{v}-\{\gamma_1\}$, we infer that $S_{vv'}$, $S_{vv''}$ and $S_{vv'''}$ contain at least one color from $F_v$. Thus we have $\parallel S'_v \parallel \le \parallel S_v \parallel - 3 \le (d(v_2)-1) + (d(v_3)-1) + (d(v_4)-1) + (d(v_5)-1) -3 \le 6 + 10 + \Delta-1 + \Delta-1 -3 = 2\Delta +11$.
\end{proof}

\begin{clm}
\label{clm:clm17}
With respect to $c'$, there exists at least one color $\alpha \in C-(F_v\cup F_{v_1})$ with multiplicity at most  one in $S'_v$.
\end{clm}
\begin{proof}
Since $v$ belongs to either configuration $A3$ or configuration $A4$, we have $\vert F_v \cup F_{v_1} \vert \le 9-3=6$. Thus $\vert C-(F_v \cup F_{v_1}) \vert \le \Delta+6$. By $Claim$ \ref{clm:clm16} we have $\parallel S'_v \parallel \le 2\Delta +11$ and from this it is easy to see that there exists at least one color $\alpha \in C-(F_v\cup F_{v_1})$ with multiplicity at most one in $S'_v$.
\end{proof}

Note that $\alpha \in C-(F_v\cup F_{v_1})$, where $\alpha$ is the color  from $Claim$ \ref{clm:clm17} is a candidate color for the edge $vv_1$. If it is not valid then there has to be a $(\theta,\alpha,vv_1)$ critical path, where $\theta \in \{1,2,3\}$. By $Claim$ \ref{clm:clm17}, $\alpha$ can be present in at most one of $S_{vv'}$, $S_{vv''}$ and $S_{vv'''}$. Without loss of generality let $\alpha \in S_{vv''}$. Thus there exists a $(2,\alpha,vv_1)$ critical path with respect to the coloring $c'$. Recolor the edge $vv'$ using the color $\alpha$ to get a coloring $c$. Clearly the recoloring is proper since $\alpha \notin S_{vv'}$ and $\alpha \notin F_v$. The recoloring is valid since if a bichromatic cycle gets formed then it has to contain the color $\alpha$ as well as a color $\gamma \in F_v(c)-\{\alpha\}$. If $\gamma = c(v,w)$, then $\alpha \in S_{vw}$, for the $(\alpha,\gamma)$ bichromatic cycle to get formed. But $v''$ is the only vertex in $N_{G'}(v)$ such that $\alpha \in S_{vv''}$. Thus $w=v''$, $\gamma=2$ and it has to be a $(\alpha,2)$ bichromatic cycle. This means that there existed a $(2,\alpha,vv')$ critical path with respect to the coloring $c'$, a contradiction by $Fact$ \ref{fact:fact1} since there already existed a $(2,\alpha,vv_1)$ critical path with respect to the coloring $c'$. Thus the coloring $c$ is  acyclic. This reduces the situation to case 1.

~~~~~

\subsection*{case 3: $\vert F_v \cap F_{v_1} \vert= 4$}

Note that in this case $\vert F_v \vert \ge 4$ and since $d(v) \le 5$, we have $d(v) = 5$. In other words, $v$ belongs to configuration $A4$. Let $S'_v$ be a multiset defined by $S'_v = S_v \setminus (F_{v_1} \cup F_{v})$. Also let $c(v,v_2)=1$, $c(v,v_3)=2$, $c(v,v_4)=3$ and $c(v,v_5)=4$.

Now try to recolor an edge incident on $v$ with a candidate color from $C-(F_v \cup F_{v_1})$. If the recoloring is valid then the situation reduces to case 2. Otherwise there has to be a bichromatic cycle created due to recoloring with one of the colors from $F_v$. This implies that $F_v \cap S'_v \neq \emptyset$. Thus we have $\parallel S'_v \parallel \le \parallel S_v \parallel - 1 \le (d(v_2)-1) + (d(v_3)-1) + (d(v_4)-1) + (d(v_5)-1) \le 6 + 10 + \Delta-1 + \Delta-1 -1 = 2\Delta +13$. Now since there are $\vert C-(F_v \cup F_{v_1})\vert \ge \Delta +12 - (4+5-4)=\Delta+7$ candidate colors and $\parallel S'_v \parallel \le 2\Delta +13$, it is easy to see that there exists at least one candidate color $\alpha$ with multiplicity at most one in $S'_v$.

Note that $\alpha \in C-(F_v\cup F_{v_1})$ is a candidate color for the edge $vv_1$. If it is not valid then there has to be a $(\theta,\alpha,vv_1)$ critical path, where $\theta \in \{1,2,3,4\}$. We know that $\alpha$ can be present in at most one of $S_{vv_2}$, $S_{vv_3}$, $S_{vv_4}$ and $S_{vv_5}$. Without loss of generality let $\alpha \in S_{vv_3}$. Thus there exists a $(2,\alpha,vv_1)$ critical path with respect to the coloring $c'$. Recolor the edge $vv_2$ using the color $\alpha$ to get a coloring $c$. Clearly the recoloring is proper since $\alpha \notin S_{vv_2}$ and $\alpha \notin F_v$. The recoloring is valid since if a bichromatic cycle gets formed then it has to contain the color $\alpha$ as well as a color $\gamma \in F_v(c)-\{\alpha\}$. If $\gamma = c(v,w)$, then $\alpha \in S_{vw}$, for the $(\alpha,\gamma)$ bichromatic cycle to get formed. But $v_3$ is the only vertex in $N_{G'}(v)$ such that $\alpha \in S_{vv_3}$. Thus $w=v_3$, $\gamma=2$ and it has to be a $(\alpha,2)$ bichromatic cycle. This means that there existed a $(2,\alpha,vv_2)$ critical path with respect to the coloring $c'$, a contradiction by $Fact$ \ref{fact:fact1} since there already existed a $(2,\alpha,vv_1)$ critical path with respect to the coloring $c'$. Thus the coloring $c$ is  acyclic. This reduces the situation to case 2.

\subsection{There exists no vertex $v$ that belongs to configuration $A2$,$A3$ or $A4$}

Then clearly by $Lemma$ \ref{lem:lem4}, we can assume that there is a vertex $v$ that belongs to configuration $A1$, i.e., $d(v)=2$. Now delete all the degree $2$ vertices from $G$ to get a graph $H$. Now since the graph $H$ is also planar, there exists a vertex $v'$ in $H$ such that $v'$ belongs to one of the configurations $A1-A4$, say $A'$. The vertex $v'$ was not already in configuration $A'$ in $G$. This means that the degree of at least one of the vertices of the configuration $A'$ i.e., $\{v'\} \cup N_{H}(v')$, got decreased by the removal of 2-degree vertices. Let $P=\{x \in \{v'\} \cup N_{H}(v'): d_{H}(x) < d_{G}(x)\}$. Let $u$ be the minimum degree vertex in $P$ in the graph $H$. Now it is easy to see that $d_{H}(u) \le 11$ since $v'$ did not belong to $A'$ in $G$.

Let $N'(u)=\{x \vert x \in N_{G}(u) $ and $ d_{G}(u)=2\}$. Let $N''(u) = N_{G}(u)-N'(u)$. It is obvious that $N''(u) = N_{H}(u)$.

Since $u \in P$ and $d_{H}(u) \le 11$, we have $\vert N'(u) \vert \ge 1$ and $N''(u) \le 11$. In $G$ let $u' \in N'(u)$ be a two degree neighbour of $u$ such that $N(u')=\{u,u''\}$. Now by induction $G-\{uu'\}$ is acyclically edge colorable using $\Delta+12$ colors. Let $c'$ be such a coloring. With respect to a partial coloring $c'$ let $F'_{u}(c')=\{c'(u,x) \vert x \in N'(u)\}$ and $F''_{u}(c')=\{c'(u,x) \vert x \in N''(u)\}$. Now if $c(u',u'') \notin F_u$ we are done since $\vert F_u \cup F_{u'} \vert \le \Delta$ and thus there are at least $12$ candidate colors which are also valid by $Lemma$ \ref{lem:lem1}.

We know that $\vert F''_v \vert \le 11$. If $c'(u',u'') \in F'_v$, then let $c=c'$. Else if $c'(u',u'') \in F''_v$, then recolor edge $u'u''$ using a color from $C-(S_{u'u''} \cup F''_v)$ to get a coloring $c$ (Note that $\vert C-(S_{u'u''} \cup F''_v) \vert \ge \Delta+12-(\Delta-1+11) = 2$ and since $u'$ is a pendant vertex in $G-\{uu'\}$ the recoloring is valid). Now if $c(u',u'') \notin F_u$ the proof is already discussed. Thus $c(u',u'') \in F'_u$.

With respect to coloring $c$, let $a \in N'(v)$ be such that $c(v,a)=c(u',u'')=1$. Now if none of the candidate colors in $C-(F_u \cup F_{u'})$ are valid for the edge $uu'$, then by $Fact$ \ref{fact:fact2}, for each $\gamma \in C-(F_u \cup F_{u'})$, there exists a $(1,\gamma,uu')$ critical path. Since $c'(v,a)=1$, we have all the critical paths passing through the vertex $a$ and hence $S_{va} \subseteq C-(F_u \cup F_{u'})$. This implies that $\vert S_{va} \vert \ge \vert C-(F_u \cup F_{u'}) \vert \ge \Delta+12-(1+ \Delta-1 -1) = 13$, a contradiction since $\vert S_{va} \vert = 1$. Thus we have a valid color for the edge $uu'$, a contradiction to the assumption that $G$ is a counter example.

\end{proof}


\begin{thebibliography}{10}

\bibitem{AMR}
{\sc N.~Alon, C.~J.~H. McDiarmid, and B.~A. Reed}, {\em Acyclic coloring of
  graphs}, Random Structures and Algorithms, 2 (1991), pp.~343--365.

\bibitem{ASZ}
{\sc N.~Alon, B.~Sudakov, and A.~Zaks}, {\em Acyclic edge-colorings of graphs},
  Journal of Graph Theory, 37 (2001), pp.~157--167.

\bibitem{AZ}
{\sc N.~Alon and A.~Zaks}, {\em Algorithmic aspects of acyclic edge colorings},
  Algorithmica, 32 (2002), pp.~611--614.

\bibitem{ART}
{\sc D.~Amar, A.~Raspaud, and O.~Togni}, {\em All to all wavelength routing in
  all-optical compounded networks}, Discrete Mathematics, 235 (2001),
  pp.~353--363.

\bibitem{MBSC4}
{\sc M.~Basavaraju and S.~Chandran}, {\em Acyclic edge coloring of 2-degenerate
  graphs}.
\newblock Submitted. Available at http://arxiv.org/abs/0803.2433v1, 2008.

\bibitem{MBSC2}
\leavevmode\vrule height 2pt depth -1.6pt width 23pt, {\em Acyclic edge
  coloring of graphs with maximum degree 4}, Journal of Graph
  Theory, 61 (2009), pp.~192--209.

\bibitem{Boro}
{\sc O.~V. Borodin}, {\em Acyclic colorings of planar graphs}, Discrete
  Mathematics, 25 (1979), pp.~211--236.

\bibitem{Burn}
{\sc M.~I. Burnstein}, {\em Every 4-valent graph has an acyclic five-coloring},
  Soobs\v c\v . Akad. Nauk Gruzin. SSR, 93 (1979).

\bibitem{Diest}
{\sc R.~Diestel}, {\em Graph Theory}, vol.~173, Springer Verlag, New York,
  2~ed., 2000.

\bibitem{Fiam}
{\sc J.~Fiamcik}, {\em The acyclic chromatic class of a graph (russian)}, Math.
  Slovaca, 28 (1978), pp.~139--145.

\bibitem{Fiam2}
\leavevmode\vrule height 2pt depth -1.6pt width 23pt, {\em Acyclic chromatic
  index of a graph with maximum valency three(russian)}, Archivum Mathematicum,
  16 (1980), pp.~81--87.

\bibitem{Fiam1}
\leavevmode\vrule height 2pt depth -1.6pt width 23pt, {\em Acyclic chromatic
  index of a subdivided graph(russian)}, Archivum Mathematicum, 20 (1984),
  pp.~69--82.

\bibitem{AFMB}
{\sc A.~Fiedorowicz and M.~Borowiecki}, {\em About acyclic edge colouring of
  planar graphs without short cycles}, Discrete Mathematics,
  doi:10.1016/j.disc.2009.06.007 (2009).

\bibitem{AMN}
{\sc A.~Fiedorowicz, M.~Hauszczak, and N.~Narayanan}, {\em About acyclic edge
  colouring of planar graphs}, Information Processing Letters, 108 (2008),
  pp.~412--417.

\bibitem{GeRa}
{\sc S.~Gerke and M.~Raemy}, {\em Generalised acyclic edge colourings of graphs
  with large girth}, Discrete Mathematics, 307 (2007), pp.~1668--1671.

\bibitem{GrePi}
{\sc C.~Greenhill and O.~Pikhurko}, {\em Bounds on the generalised acyclic
  chromatic numbers of bounded degree graphs}, Graphs and Combinatorics, 21
  (2005), pp.~407--419.

\bibitem{Grun}
{\sc B.~Gr\"unbaum}, {\em Acyclic colorings of planar graphs}, Israel Journal
  of Mathematics, 14 (1973), pp.~390--408.

\bibitem{JWu1}
{\sc J.~Hou, J.~Wu, G.~Liu, and B.~Liu}, {\em Acyclic edge colorings of planar
  graphs and series-parallel graphs}, Science in China Series A: Mathematics,
  52 (2009), pp.~605--616.

\bibitem{KSZ}
{\sc A.~Kostochka, E.~Sopena, and X.~Zhu}, {\em Acyclic and oriented chromatic
  numbers of graphs}, J. Graph Theory, 24 (1997), pp.~331--340.

\bibitem{MolReed}
{\sc M.~Molloy and B.~Reed}, {\em Further algorithmic aspects of lov\'asz local
  lemma}, in Proceedings of the 30th Annual ACM Symposium on Theory of
  Computing, 1998, pp.~524--529.

\bibitem{MNS1}
{\sc R.~Muthu, N.~Narayanan, and C.~R. Subramanian}, {\em Improved bounds on
  acyclic edge coloring}, Electronic notes in discrete mathematics, 19 (2005),
  pp.~171--177.

\bibitem{MNS2}
\leavevmode\vrule height 2pt depth -1.6pt width 23pt, {\em Optimal acyclic edge
  coloring of grid like graphs}, in Proceedings of the 12th International
  Conference, COCOON, LNCS 4112, 2006, pp.~360--367.

\bibitem{NesWorm}
{\sc J.~N\v{e}set\v{r}il and N.~C. Wormald}, {\em The acyclic edge chromatic
  number of a random d-regular graph is d+1}, Journal of Graph Theory, 49
  (2005), pp.~69--74.

\bibitem{Salavatipour}
{\sc M.~R. Salavatipour}, {\em Graph Coloring via the Discharging Method}, PhD
  thesis, Dept. of Computer Science, University of Toronto, 2003.

\bibitem{Skul}
{\sc S.~Skulrattankulchai}, {\em Acyclic colorings of subcubic graphs},
  Information processing letters, 92 (2004), pp.~161--167.

\bibitem{HeuvelMcG}
{\sc J.~van~den Heuvel and S.~McGuinness}, {\em Coloring the square of a planar
  graph}, Journal of Graph Theory, 42 (2003), pp.~110--124.

\end{thebibliography}

\section{Appendix}

\textbf{These proofs are for the convenience of the referee. This section will be deleted in the final version.}

\begin{lem1}
\cite{MBSC2}
A candidate color for an edge $e=uv$, is valid if $(F_u \cap F_v) - \{c(u,v)\} = (S_{uv} \cap S_{vu}) = \emptyset$.
\end{lem1}
\begin{proof}
Any cycle containing the edge $uv$ will also contain an edge incident on $u$ (other than $uv$) as well as an edge incident on $v$ (other than $uv$). Clearly these two edges are colored differently since $(S_{uv} \cap S_{vu}) = \emptyset$. Thus the cycle will have at least 3 colors and therefore any of the candidate colors for the edge $uv$ is valid.
\end{proof}

\begin{lem1}
\cite{MBSC2}, \cite{MBSC4}
Let $u,i,j,a,b \in V(G)$, $ui,uj,ab \in E$. Also let $\{\lambda,\xi\} \in C$ such that $\{\lambda,\xi\} \cap \{c(u,i),c(u,j)\} \neq \emptyset$ and $\{i,j\} \cap \{a,b\} = \emptyset$. Suppose there exists an ($\lambda$,$\xi$,$ab$)-critical path that contains vertex $u$, with respect to a valid partial coloring $c$ of $G$. Let $c'$ be the partial coloring obtained from $c$ by the color exchange with respect to the edges $ui$ and $uj$. If $c'$ is proper, then there will not be any ($\lambda$,$\xi$,$ab$)-critical path in $G$ with respect to the partial coloring $c'$.
\end{lem1}
\begin{proof}
Firstly, $\{\lambda,\xi\} \neq \{c(u,i),c(u,j)\}$. This is because, if there is a ($\lambda$,$\xi$,$ab$)-critical path that contains vertex $u$, with respect to a valid partial coloring $c$ of $G$, then it has to contain the edge $ui$ and $uj$. Since $i \notin \{a,b\}$, vertex $i$ is an internal vertex of the critical path which implies that both the colors $\lambda$ and $\xi$ (that is $c(u,i)$ and $c(u,j)$) are present at vertex $i$. That means $c(u,j) \in S_{ui}$ and this contradicts $Fact$ \ref{fact:fact3}, since we are assuming that the color exchange is proper. Thus $\{\lambda,\xi\} \neq \{c(u,i),c(u,j)\}$.

Now let $P$ be the ($\lambda$,$\xi$,$ab$) critical path with respect to the coloring $c$. Without loss of generality assume that $\gamma = c(u,i) \in \{\lambda,\xi\}$. Since vertex $u$ is contained in path $P$, by the maximality of the path $P$, it should contain the edge $ui$ since $c(u,i)=\gamma \in \{\lambda,\xi\}$. Let us assume without loss of generality that path $P$ starts at vertex $a$ and reaches vertex $i$ before it reaches vertex $u$. Now after the color exchange with respect to the edges $ui$ and $uj$, i.e., with respect to the coloring $c'$, there will not be any edge adjacent to vertex $i$ that is colored $\gamma$. So if any ($\lambda$,$\xi$) maximal bichromatic path starts at vertex $a$, then it has to end at vertex $i$. Since $i \neq b$, by $Fact$ \ref{fact:fact1} we infer that the ($\lambda$,$\xi$,$ab$) critical path does not exist.

\end{proof}

\end{document}